\documentclass[longbibliography, aps, prl, amsmath, amssymb,
amsfonts, twocolumn,superscriptaddress, footinbib]{revtex4-2}
\usepackage[german,american,english]{babel}
\usepackage{graphicx}
\usepackage{dcolumn}
\usepackage{bm}
\usepackage{amssymb}
\usepackage{amsmath}
\usepackage{amsfonts}
\usepackage{epsfig}
\usepackage{mathbbol}
\usepackage{latexsym}
\usepackage{theorem}
\usepackage{dcolumn}
\usepackage{algorithm}
\usepackage{algpseudocode}
\usepackage{hyperref}
\usepackage{float}
\usepackage[normalem]{ulem}
\usepackage{xcolor}
\usepackage{epstopdf}
\usepackage{physics}
\usepackage{lipsum}
\usepackage{microtype}
\hypersetup{                    
	colorlinks,
	citecolor=blue,
	filecolor=blue,
	linkcolor=blue,
	urlcolor=blue
}
\usepackage[utf8]{inputenc}
\UseRawInputEncoding

\def \blue #1 {\textcolor{blue}{#1}}

\begin{document}
	
	\title{Dual-Capability Machine Learning Models for Quantum Hamiltonian Parameter Estimation and Dynamics Prediction}
	
	\author{Zheng An}
	
	\affiliation{Department of Physics, The Hong Kong University of Science and Technology, Clear Water Bay, Kowloon, Hong Kong, China}

	\author{Jiahui Wu}
	
	\affiliation{Department of Physics, The Hong Kong University of Science and Technology, Clear Water Bay, Kowloon, Hong Kong, China}
 
	\author{Zidong Lin}
	
	\affiliation{Shenzhen SpinQ Technology Co., Ltd., 518043, Shenzhen, China}

        \author{Xiaobo Yang}
        
        \affiliation{Shenzhen SpinQ Technology Co., Ltd., 518043, Shenzhen, China}

        \author{Keren Li}

        \email{likr@szu.edu.cn}
        
        \affiliation{College of Physics and Optoelectronic Engineering, Shenzhen University, Shenzhen 518060, China}
        
        \affiliation{Quantum Science Center of Guangdong-Hong Kong-Macao Greater Bay Area (Guangdong), Shenzhen 518045. China}
	
	\author{Bei Zeng}%
	
	\email{zengb@ust.hk}
	
	\affiliation{Department of Physics, The Hong Kong University of Science and Technology, Clear Water Bay, Kowloon, Hong Kong, China}
	\date{\today}
	
	\begin{abstract}  
	Recent advancements in quantum hardware and classical computing simulations have significantly enhanced the accessibility of quantum system data, leading to an increased demand for precise descriptions and predictions of these systems. Accurate prediction of quantum Hamiltonian dynamics and identification of Hamiltonian parameters are crucial for advancements in quantum simulations, error correction, and control protocols. This study introduces a machine learning model with dual capabilities: it can deduce time-dependent Hamiltonian parameters from observed changes in local observables within quantum many-body systems, and it can predict the evolution of these observables based on Hamiltonian parameters. Our model's validity was confirmed through theoretical simulations across various scenarios and further validated by two experiments. Initially, the model was applied to a Nuclear Magnetic Resonance quantum computer, where it accurately predicted the dynamics of local observables. The model was then tested on a superconducting quantum computer with initially unknown Hamiltonian parameters, successfully inferring them. 
    Our approach aims to enhance various quantum computing tasks, including parameter estimation, noise characterization, feedback processes, and quantum control optimization.
	\end{abstract}

	\maketitle 
	
	\emph{Introduction --- } The quest to precisely predict and interpret the dynamics of quantum many-body systems is crucial for advancing quantum technology applications~\cite{Gebhart2023}. These efforts encounter formidable challenges, largely attributable to the inherent properties of quantum systems, such as quantum states, which are characterized by rapidly escalating complexity~\cite{Leimkuhler2004}. Moreover, Hamiltonians are fundamental to the dynamics of quantum many-body systems, and discerning the parameters of Hamiltonians from quantum system dynamics is a quintessential issue in quantum computing~\cite{mermin2007quantum}, quantum control~\cite{Dong2010,PRXQuantum.3.020357}, and quantum simulations~\cite{RevModPhys.86.153}.
	
	Predicting quantum state dynamics is challenging due to the exponential scale of quantum state space, and various methodologies have been explored to address this complexity~\cite{huang2022learning,Caro2023,PhysRevX.10.011006,Carleo2017,Mohseni2022deeplearningof,sornsaeng2023quantum}. Even modestly sized quantum systems can have state spaces larger than the number of atoms in the universe, making direct computation often intractable~\cite{Calabrese2005,jerbi2023power}. Quantum state evolution, governed by Hamiltonians as described by the Schr{\"o}dinger or Lindblad equations for environmental interactions~\cite{PhysRevLett.79.3101, I.S.A.R.1994}, presents significant computational challenges. Inversely, learning Hamiltonians from observed state dynamics involves reverse engineering the system's behavior, a task complicated by the complex, non-classical behaviors of quantum systems and the vast disparity between the sizes of quantum and observable state spaces~\cite{PhysRevA.77.032322,PhysRevB.95.041101,PhysRevLett.118.020401}. The unitary nature of quantum evolution, which preserves initial state information, further complicates this task.

	Machine learning (ML), particularly deep learning, excels at interpreting high-dimensional data and recognizing complex patterns, making it an invaluable tool for quantum system analysis. Deep learning has notably advanced quantum sciences in areas such as learning phase transitions~\cite{PhysRevB.97.045207,Carrasquilla2017,Nieuwenburg2017,an2021learning}, quantum state tomography~\cite{PhysRevX.8.011006,Torlai2018,Wang2022,an2023unified}, optimizing quantum circuits~\cite{Foesel2021,Ostaszewski2021,cao2022}, solving quantum control issues~\cite{Bukov2018,An2019,An2021}, and exploring quantum many-body dynamics~\cite{Carleo2017,Mohseni2022deeplearningof,Gao2017,PhysRevLett.125.100503,PhysRevX.10.011006,Lin2022,sornsaeng2023quantum}. However, are there any models capable of simultaneously addressing multiple demands?

	\begin{figure}[!htbp]
		\centering
		\includegraphics[width=\columnwidth]{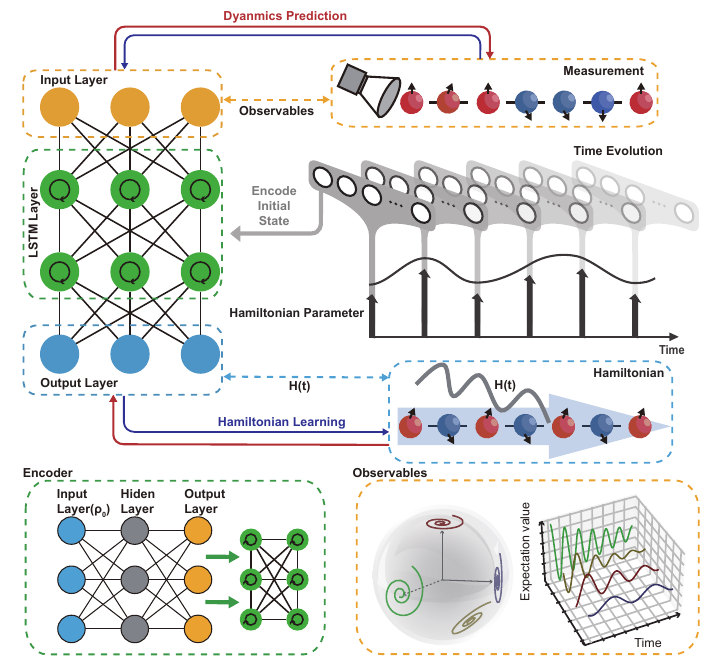}
		\caption{Schematic Representation of the Machine Learning Model for Learning Hamiltonian Parameters and Predicting Quantum Dynamics. \textbf{Top right}: An isolated spin system is initialized in a product state and then evolved under the Hamiltonian $H(t)$. Measurement sequence showing the dynamic state of quantum observables over time, captured through periodic assessments. Hamiltonian learning depicted through time evolution, where the model infers time-dependent Hamiltonian $H(t)$ from the dynamics of observables. Dynamics prediction depicted through Hamiltonian, where the model forecasts the time-dependent behavior of observables based on the input Hamiltonian parameters. \textbf{Top left}: Detailed architecture of the LSTM neural network, highlighting the input, LSTM, and output layers. 
        \textbf{Bottom left}: The initial states expectation values are encoded via an encoder into the initial hidden state of the LSTM. \textbf{Bottom right}: A Bloch sphere representation illustrates the trajectory of the local spins, allowing us to determine their expectation values as functions of time.}
		\label{fig:schemit}
	\end{figure}

     \begin{figure*}[ht]
      \begin{center}
      \includegraphics[width=\textwidth]{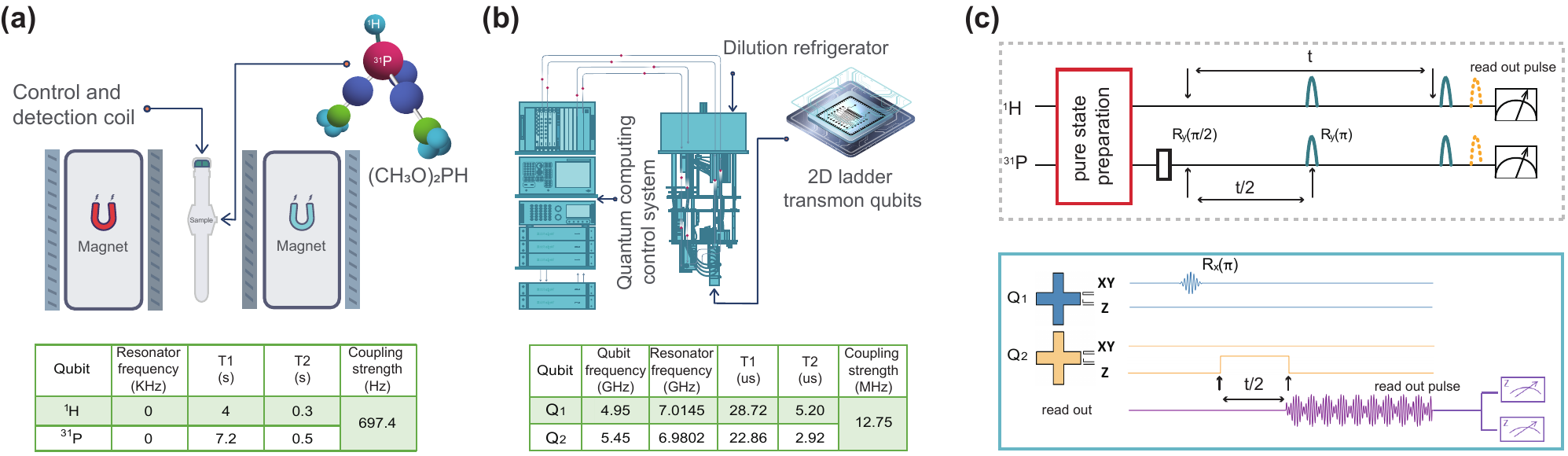}
        \caption{(a) Molecular structure of Dimethylphosphite ((CH$_3$O)$_2$PH): $^1$H and $^{31}$P serve as the two qubits for NMR quantum computer. The specified parameters are listed in the accompanying table. (b) Two-qubit superconducting quantum device, where qubits with frequencies of $4.95$GHz and $5.45$GHz are employed as two qubits. The specified parameters are listed in the accompanying table. (c) lists the quantum circuits, where circuit in dashed (solid) line box is for NMR (superconducting qubits) experiments.}
        \label{fig2}
        \end{center}
    \end{figure*}
    
	Here, we present a methodology that exploits recurrent neural networks (RNNs) for two tasks: the prediction of quantum dynamics and the elucidation of quantum Hamiltonians. Our model is based on the ``Long Short-Term Memory" (LSTM)~\cite{Hochreiter1997}, a well-recognized type of RNN, but
	with some notable differences. We add an encoding network used to encode the initial hidden variables of the LSTM, allowing our model to be applied to arbitrary initial states. This flexible architecture lets us freely switch the input data and the labeled data, enabling our model to learn bidirectionally. In addition, we combine the two tasks into the same set of training data, making our model capable of bidirectional learning (see Fig.~\ref{fig:schemit}). This bidirectionality not only augments the model's analytical breadth but also underscores its adaptability for comprehensive quantum system investigations. To demonstrate the capability of our model, we analyze data collected from experiments conducted on both Nuclear Magnetic Resonance (NMR) and superconducting quantum computers (see Fig.~\ref{fig2}). These experiments are designed to predict the evolution of a quantum system and to infer parameters of a system's Hamiltonian, showcasing the effectiveness in diverse quantum environments.
	
	\emph{Methodology --- }The fundamental concepts of quantum dynamics prediction and Hamiltonian learning are framed within the context of an $n$-qubit Hermitian Hamiltonian, expressed as $H=\sum_{a=1}^m \lambda_a E_a$. Here, $\lambda_a$ are real coefficients, and $E_a$ represents local Pauli operators affecting at most $k$ particles, with $m=O(\text{poly}(n))$. 

    In the domain of predicting quantum Hamiltonian dynamics, unitary evolution can be represented using 
    the Hamiltonian as $U\left(t, t_0\right)=\exp \left(-\frac{2 \pi i}{\hbar} H\left(t-t_0\right)\right)$, determined by a set of adjustable parameters $\{\lambda_a\}$. 
    This expression describes any unitary transformation from an initial time $t_0$ to a later time $t$ using an exponential function that involves the integral over $H$. 

    Thus, predicting quantum dynamics is to utilize measurements $\{O(0) \dots O(t_0)\}$ from initial states to project the state of the system at a future finite time, thereby elucidating the evolutionary operator $U(t)$. 
    In the domain of Hamiltonian learning, the emphasis is on reconstructing $H$ from observed quantum dynamics. Unlike traditional methods that require preparing specific quantum states, this methodology focuses on analyzing the temporal evolution of $U(t)$. Conclusively, Hamiltonian learning is to learn the dynamic $\lambda_a(t)$ of Hamiltonian $H(t)$, based on the measurements observed.
	
	We use a supervised learning approach for two problems: predicting the time-dependent expectation values of observables $O(t)$ based on the time-dependent Hamiltonian, and inferring the Hamiltonian parameters $\lambda_a(t)$ from these dynamics. Fig.~\ref{fig:schemit} illustrates the main concept of this methodology. The model is trained on varied trajectories $H(t)$, sampled from a broad range of functions, enabling it to predict the dynamics of $O(t)$ for new, unseen trajectories. We found employing random Gaussian processes for sampling these trajectories particularly effective, as they allow flexible data fitting by controlling mean and variance. Specifically, our approach incorporates a mixture of Gaussian processes~\cite{Liu2019} with varied correlation times. This method, through the use of Gaussian random functions generated by eigendecomposition of correlation matrices, has proven efficient in other areas, such as predicting band structures in topological materials~\cite{PhysRevX.11.021052}. Details and specific parameters are provided in the supplementary material~\cite{supp}.
	
	\emph{Numerical Validation --- }We begin by examining the one-dimensional Transverse Field Ising Model (TFIM):
	\begin{equation}
		H_{\text{TFIM}} = -\sum_{i=0}^{N-1}(  J\sigma_i^z \sigma_{i+1}^z+B(t)\sigma_i^x ).
	\end{equation}
    Here, $\sigma_i^\alpha$ denotes Pauli operators at site $i$, where $\alpha$ can be $x, y,$ or $z$. We fix $J=1$ and focus on the external driving field $B(t)$.
	
	Next, we apply our model to predict quantum dynamics and learn Hamiltonian parameters under varying conditions. These scenarios include different driving fields and quantum many-body systems affected by decoherence or functioning in a noiseless environment. Further training and evaluation details are available in the supplementary material~\cite{supp}.
	
	\begin{figure}[!htbp]
		\centering
		\includegraphics[width=\columnwidth]{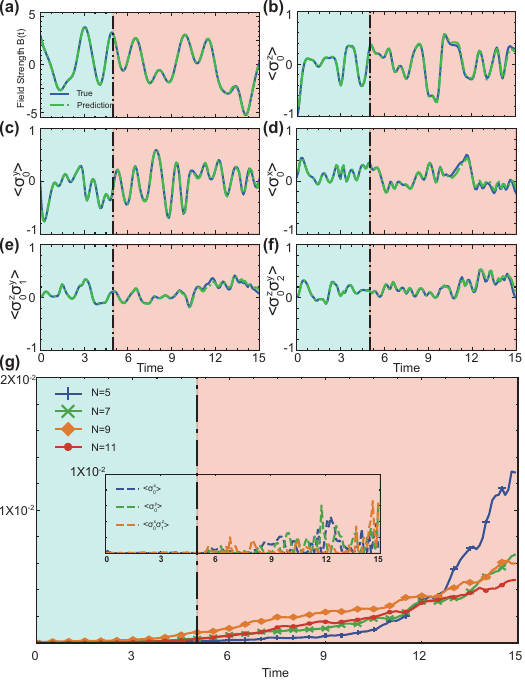}
		\caption{Overview of our model's predictive accuracy on the noiseless dynamics of 1D TFIM dynamics under arbitrary driving field. Initially trained within $t\in [0,5]$, the model's performance extends across a broader time span ($t\in [0,15]$) and various driving conditions. Displayed results include: (a) actual vs. predicted external driving fields $B(t)$, (b-f) local observables $\langle \sigma_0^\alpha \rangle$ ($\alpha = x,y,z$) and $\langle \sigma_0^z \sigma_{0+l}^y \rangle$ ($l=1,2$) in a size $N=5$ transverse-field Ising spin ring, and (g) prediction loss, measured as MSE, averaged over each 100 instances with varying system sizes ($N=5,7,9,11$). Inset: Squared Error Analysis for a Sample Random Field at N = 5 with three local observables: $\langle \sigma_0^x \rangle$, $\langle \sigma_0^y \rangle$, and $\langle \sigma_0^x \sigma_{1}^z \rangle$. The model evaluated spins initialized in a randomly directed, translationally-invariant product state, with predictions beyond the training range highlighted in light orange for $t\in [5,15]$.
		}
		\label{fig:results5}
	\end{figure}

	Fig.~\ref{fig:results5} illustrates the alignment between our model's predictions and the ground truth under unseen random driving fields, maintaining a high accuracy level over an extended time interval ($t\in [0,15]$). Fig.~\ref{fig:results5}(a-f) confirm that the model reliably predicts outcomes under such conditions. By employing an approach that encompasses both dynamics prediction and Hamiltonian learning, the model accurately predicts selected local observables $\langle \sigma_0^\alpha \rangle$ ($\alpha = x,y,z$) and $\langle \sigma_0^z \sigma_{0+l}^y \rangle$ ($l=1,2$) based on the external field trajectory (see Fig.~\ref{fig:results5}(b-f)), and deduces the external driving field $B(t)$ from the dynamics of these observables (see Fig.~\ref{fig:results5}(a)). Our models were evaluated by training on noiseless evolutions across systems of different sizes ($N=5, 7, 9, 11$). Fig.~\ref{fig:results5}(g) shows that the Mean Squared Error (MSE) losses for all observables stayed below $10^{-3}$ during the initial training period ($t\in [0,5]$). In the extrapolation period ($t\in (5,15]$), although losses increased, peaking at around $10^{-2}$, they remained relatively low. As illustrated in the inset, the fluctuations in the loss tend to increase over time, leading to a statistical rise in the overall loss. However, the errors associated with individual local observables remain small ($1\%$). Further exploration into different field types and noise scenarios is detailed in the supplementary material~\cite{supp}.

    \emph{Experimental Validation --- }To further validate our model, we test it using experimental data.
    To leverage the capability of our model in predicting quantum dynamics, we employed a desktop NMR quantum computer known as Gemini~\cite{hou2021spinq}. As shown in Fig.~\ref{fig2}(a), the device comprises two qubits, represented by the nuclei $^1$H and $^{31}$P in Dimethylphosphite molecules. 
    With the application of an external control field, we can align the Hamiltonian of our experimental system to match that of the theoretical two-qubit model,
    \begin{equation} \label{exp_NMR}
        H_{t} = \frac{\pi}{2}B(t)\sigma_0^z \sigma_1^z.
    \end{equation}
    where $B(t)$ is a constant as $B_0=697.4$~Hz, the J-coupling strength of molecular Hamiltonian.
    Further details about the setup are available in the supplementary material~\cite{supp,lu2017enhancing}. 
    To demonstrate that our model can reproduce the dynamics of the two-qubit system, which depend on $B(t)$, we conducted the experiment (dashed line box) as illustrated in the circuit of Fig.~\ref{fig2}(c). We collected data from these experiments and subsequently used our model to analyze and compare the results, verifying its efficacy and predictive capabilities. 
    
    The experiments were executed through the following steps:
    First, a relaxation-based method was employed to generate a pseudo-pure state, $\ket{00}$, from a thermal equilibrium state, as extensively detailed in~\cite{li2016approximation}. 
    Following this, an $R_y(\pi/2)$ rotation was applied to the second qubit, $^{31}$P, converting the system into the two-qubit state $\ket{0+}$. 
    Then, the system underwent a period of free evolution lasting for a duration $T$.
    During this evolution phase, $250$ time-points were sampled at uniform intervals of $200$us, labeled as $\Delta t$, beginning from $0$ and evenly distributed over the interval $[0, T)$, where $T=50ms$.
    Remarkably, a key aspect of the experiment was the design of a time-varying $B(t)$. 
    Given that $B_0$ in Eq.~\ref{exp_NMR} is constant,
    the $\Delta t$ for each  $m$-th step was adjusted as,
    \begin{eqnarray}
        \Delta t \rightarrow \int_{(m-1)\Delta t}^{m\Delta t} B(t) dt/B_0.
    \end{eqnarray}
    Finally, at each sampled time-point, measurements were taken for $15$ observables, comprising the set $\{\sigma_I, \sigma_x, \sigma_y, \sigma_z\}^{\otimes 2}$, excluding $\sigma_I \otimes \sigma_I$.
    
    \begin{figure}[!htbp]
                \centering
                \includegraphics[width=\columnwidth]{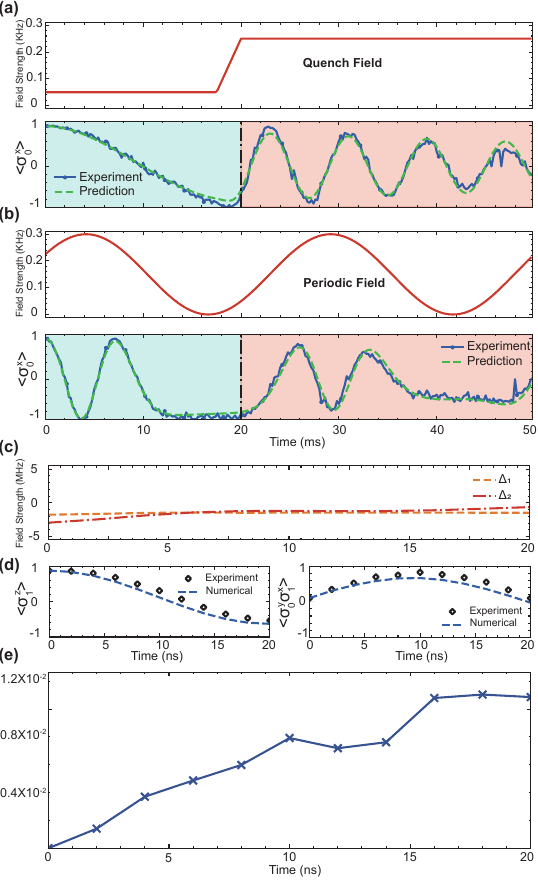}
                \caption{Experimental results of our model from two-qubit NMR and superconducting quantum computers. (a) and (b) simulate quantum system evolution under different driving fields: a quench and periodic driving, trained over the interval $t\in [0 \text{ms}, 20 \text{ms}]$ and evaluated up to $t\in [0 \text{ms}, 50 \text{ms}]$. The local observable $\langle \sigma_0^x \rangle$) is displayed. Spins were initialized in a randomly directed, translationally-invariant product state. The light orange segment from $t\in [20\text{ms}, 50\text{ms}]$ highlights predictions extending beyond the training phase. 
                (c), (d), and (e) display results that assess the capability of our model to infer parameters of an unknown Hamiltonian. Specifically, (c) presents the field strengths $\Delta 0$ and $\Delta 1$ as inferred by our model. (d) compares the expected values of local observables with the actual experimental data. (e) quantifies the prediction loss using the MSE, averaged across all observables for the two qubits.}
                \label{fig:exp}
            \end{figure}
    
    Two scenarios time varying $B(t)$ were considered: a quench and periodic driving field.
    Utilizing our trained model, we provide predictions for time-varying local observables, expressed in terms of Pauli operators $\langle \sigma_0^\alpha \rangle$ ($\alpha=x,y,z$).
    Fig.~\ref{fig:exp}(a)-(b) display these results, which indicates that our model's predictions closely align with the actual measurement trajectories from a two-qubit quantum computer, with the training conducted over a time range of $t \in [0 , \text{ms}, 20 , \text{ms}]$. This alignment showcases the model's robustness and highlights its potential for exploring complex quantum behaviors across various timescales.
    
    To assess the capability of our model in inferring parameters of an unknown Hamiltonian from observed dynamics, we conducted tests to learn detuning on a superconducting qubit system.
    Detuning is critical not only as an error source but also as a potential computational resource~\cite{ganzhorn2020benchmarking,shen2021ultrafast}. Besides, it plays a pivotal role in the development of quantum computers on superconducting platforms.
    In this section, we employ the model to infer the strength of detuning during the operation of logical quantum gates. This analysis is based on data collected from a superconducting chip, which is configured as a 2D ladder with 20 transmon qubits.
    As shown in Fig.~\ref{fig2} (b), the qubits operating at frequencies of $4.95$~GHz and $5.45$~GHz, were used and designated as $Q_1$ and $Q_2$.
    
    During this experiment, a square-shaped wave was applied to the Z bias control line of $Q_2$. The amplitude of the square-shaped wave was fine-tuned to equalize the frequencies of the two qubits. When the frequencies were synchronized, the qubits were expected to undergo a SWAP evolution, characterized by the interaction, $\sigma^x_1 \sigma^x_2 + \sigma^y_1 \sigma^y_2$. The coupling strength of this interaction is scaled by $B_0=12.75$~Mhz.
    However, inaccuracy in the pulse and external noises can introduce detuning, resulting in deviations in the expected Hamiltonian,
    \begin{eqnarray} \label{exp_SP}
        H_t=\frac{B_0}{2} (\sigma^x_1 \sigma^x_2 + \sigma^y_1 \sigma^y_2)+\Delta_1\sigma_1^z+\Delta_2\sigma_2^z.
    \end{eqnarray}
    Here $\Delta_1$ and $\Delta_2$ represent the detuning errors for $Q_1$ and $Q_2$, which may distort the logic gates designed by this method. 
    Therefore, the purpose is to determine the magnitude of the detuning ($\Delta_1$ and $\Delta_2$). To achieve this, we utilize our proposed model to analyze measurements taken from experiments on this superconducting qubit.

    As illustrated in the circuit of Fig.~\ref{fig2}(c), the experiment (solid line box) was executed through the following steps:
    First, from the ground state $\ket{00}$, the system was set to the state $\ket{10}$. This was achieved by applying a $\pi$ pulse to $Q_1$ on the $x$-axis.  
    Following this, a square-shaped wave was applied to the Z bias control line of $Q_2$. The duration of the square-shaped wave was set to $T=20$~ns, during which it was assumed that $11$ time-points were sampled at regular intervals of $2$~ns.
    Finally, at each of these points, measurements were conducted for selected observables from the set $\{\sigma_I, \sigma_x, \sigma_y, \sigma_z\}^{\otimes 2}$. This step involved performing appropriate measurement operations on the qubits and subsequently reading signals from the readout resonator of both qubits. Remarkably, quantum gates, square-shaped wave used in our experiments were calibrated according to~\cite{mckay2017efficient,neill2018blueprint,motzoi2009simple}, and we specify them in supplementary material~\cite{supp}. 
    
    Upon analyzing the data collected, we could infer the detuning strengths, labeled as $\Delta_1$ and $\Delta_2$. 
    As shown in Fig.~\ref{fig:exp}(c), the model extracts a time series of $\Delta_1$ and $\Delta_2$.
    This accuracy is evidenced by the agreement between the numerically simulated expected values of the local observables, $\langle \sigma_1^z \rangle$ and $\langle \sigma_0^y\sigma_1^x \rangle$, based on the inferred detuning strengths, and their corresponding experimental measurements. This alignment is clearly illustrated in Fig.~\ref{fig:exp}(d).
    Finally, the precision of this evaluation is quantified using the MSE function, which measures the average prediction error across all local observables for both qubits. This metric is displayed in Fig.~\ref{fig:exp}(e). The low MSE($1\%$) underscores the model's capability to reveal intricate quantum behaviors, proving its value for parameter inference in experimental settings where some conditions are unknown.
	
		\emph{Conclusion --- }In summary, we have developed a machine learning framework that utilizes LSTM to effectively predict quantum dynamics and infer Hamiltonian parameters. Our approach has been tested on both NMR and superconducting quantum computing platforms, demonstrating accuracy in predicting observable dynamics and determining unknown Hamiltonian parameters. Our model supports two functionalities: dynamics prediction and deduce Hamiltonian parameters, making it suitable for applications in quantum computing applications such as quantum simulation, parameter estimation, quantum control, and error mitigation. The empirical validation across theoretical simulations and experimental setups confirms the model's robustness and adaptability, showing its potential for future quantum computing applications. This study contributes to the ongoing efforts to leverage machine learning in understanding and manipulating complex quantum systems.
  
	\begin{acknowledgments}
	The authors thank Lingzhen Guo and Naeimeh Mohseni for helpful discussions. This work is supported by the GRF (grant no. 16305121).
        \end{acknowledgments}
	\bibliographystyle{apsrev4-2} \bibliography{main.bbl}

\pagebreak
\onecolumngrid
\begin{center}
	\textbf{\large Supplemental Material for ``Dual-Capability Machine Learning Models for Quantum Hamiltonian Parameter Estimation and Dynamics Prediction''}
\end{center}
\setcounter{equation}{0}
\setcounter{figure}{0}
\setcounter{table}{0}
\setcounter{page}{1}
\makeatletter
\renewcommand{\theequation}{S\arabic{equation}}
\renewcommand{\thefigure}{S\arabic{figure}}

		\section{The neural network architecture}
	\begin{figure}[htbp]
		\centering
		\includegraphics[width=0.48\textwidth]{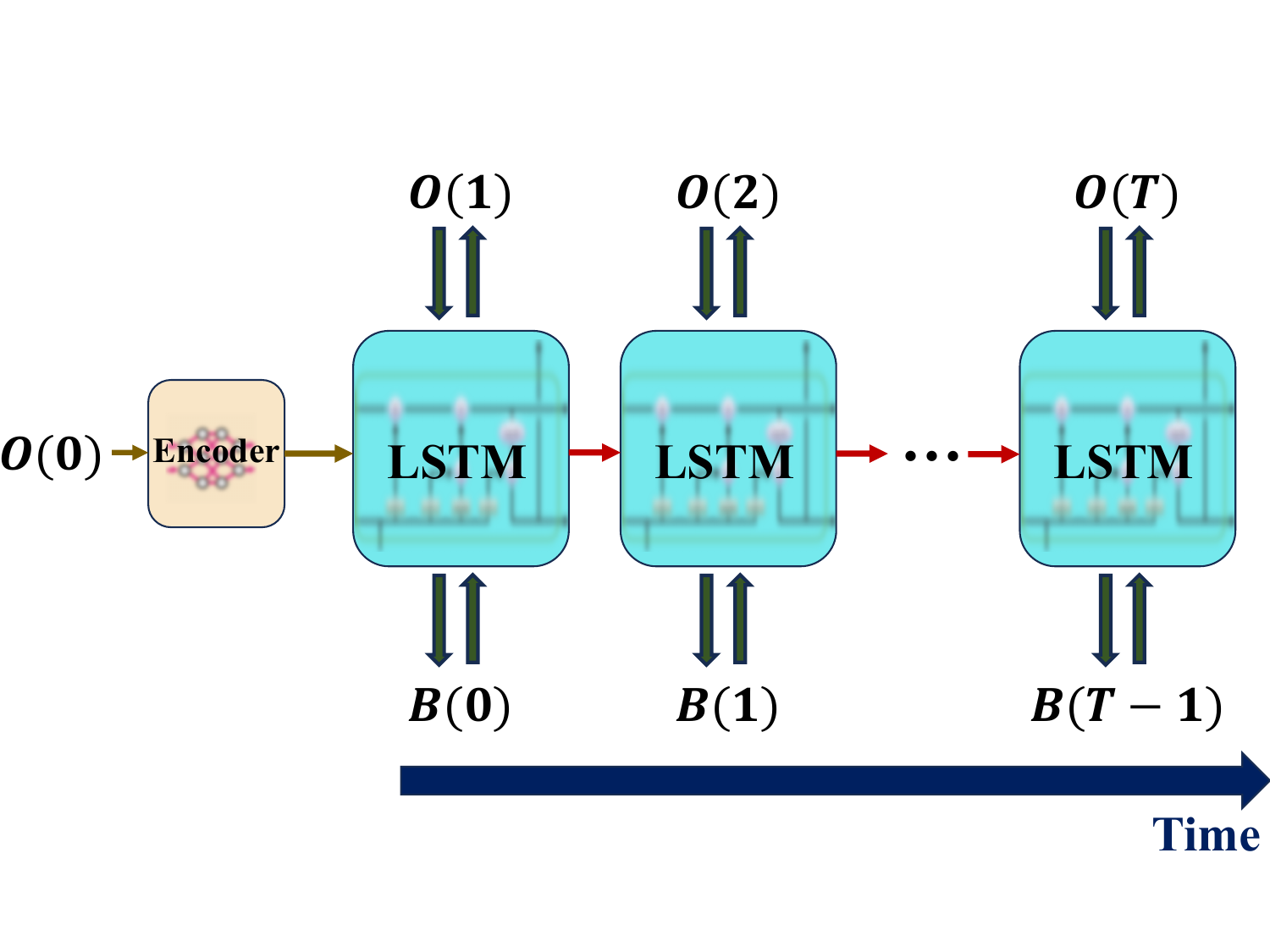}
		\caption{A schematic representation of our training strategy using a LSTM-NN. The horizontal arrows signify the internal neural memory content being transferred to the subsequent time step. The initial state's expectation values $O(0)$ are encoded with a NN as an initial hidden state of the LSTM. If the driving field $B(t)$ is input, the output will be the full evolution of the expectation values of a subset of observables of interest $O(t)$ for quantum state evolution prediction. Conversely, if the expectation values $O(t)$ are input, the output will be the driving field $B(t)$, facilitating dynamic Hamiltonian learning.}
		\label{fig:learning}
	\end{figure}
	
	For our Neural Network architecture, we utilize the "Long Short-Term Memory" (LSTM)~\cite{Hochreiter1997}, a well-recognized type of RNN. This choice is influenced by the LSTM's inherent capabilities. Like all RNNs, LSTM-NNs have a temporal structure and respect the principle of causality, making them apt for representing differential equations, often referred to as equations of motion. Uniquely, LSTMs can capture both long-term and short-term dependencies, an invaluable trait when dealing with complex non-Markovian dynamics, which involve dependencies that don't follow a simple Markov chain. Consequently, these combined attributes suggest that the LSTM architecture is highly promising for the prediction of many-body dynamics, even for researchers from diverse fields (refer to Appendix A for a concise summary of the LSTM architecture).
	
	In our model, we leverage the capabilities of LSTM (Fig.~\ref{fig:learning}). We encode the initial state of the quantum system into the LSTM's initial hidden state, allowing us to handle any initial state. This flexible architecture lets us freely transform the input data and the labeled data, enabling our model to learn bidirectionally. This streamlined approach not only simplifies the data acquisition process but also confers a significant advantage in the simultaneous training of both prediction quantum state evolution and Hamiltonian learning, thereby enhancing the efficiency and effectiveness of these interconnected methodologies in elucidating quantum states and estimating Hamiltonian parameters.
	
	In Table.~\ref{table:simplified_model}, we present the configuration of our Encoder neural network and the LSTM. The Encoder employs the hyperbolic tangent (tanh) activation function for all layers, excluding the final layer where a linear activation function is utilised. The 'adam' optimizer is consistently applied across all scenarios. For the LSTM, in addition to the driving field or observation values, the current time variable is also inputted into the model. This practice appears to improve the accuracy of the model's predictions.
	\begin{table*}[htbp]
		\centering
		\begin{tabular}{|c|c|c|c|c|}
			\hline
			\textbf{NN Type} & \textbf{Number of Layer} & \textbf{N} & \textbf{Input Size} & \textbf{Output Size}\\
			\hline
			Encoder for hidden state & 4 & 500 & 3&500\\
			Encoder for cell state & 4 & 500 & 3&500 \\
			LSTM & 4 & 500& $1+1$ or $9(l-1)+3+1$ & $9(l-1)+3$ or $1$ \\
			
			\hline
		\end{tabular}
		\caption{Architectural design of the encoder neural network and the LSTM. The term \textbf{N} represents the number of neurons per layer. The maximum distance between spins on a ring is denoted by $l$. The expression $9(l - 1) + 3$ signifies the aggregate count of all observable variables chosen for training, as discussed in the main text. These observable variables include all first and second-order moments (correlators) of spin operators. The number 3 corresponds to the parameters defining the initial state, with our initial product state being identified by the expectation values of the first-order moments of these spin operators.}
		\label{table:simplified_model}
	\end{table*}

	\section{Recurrent Neural Networks and Structure LSTM}
	
\begin{figure}[htbp]
		\centering
		\includegraphics[width=0.48\textwidth]{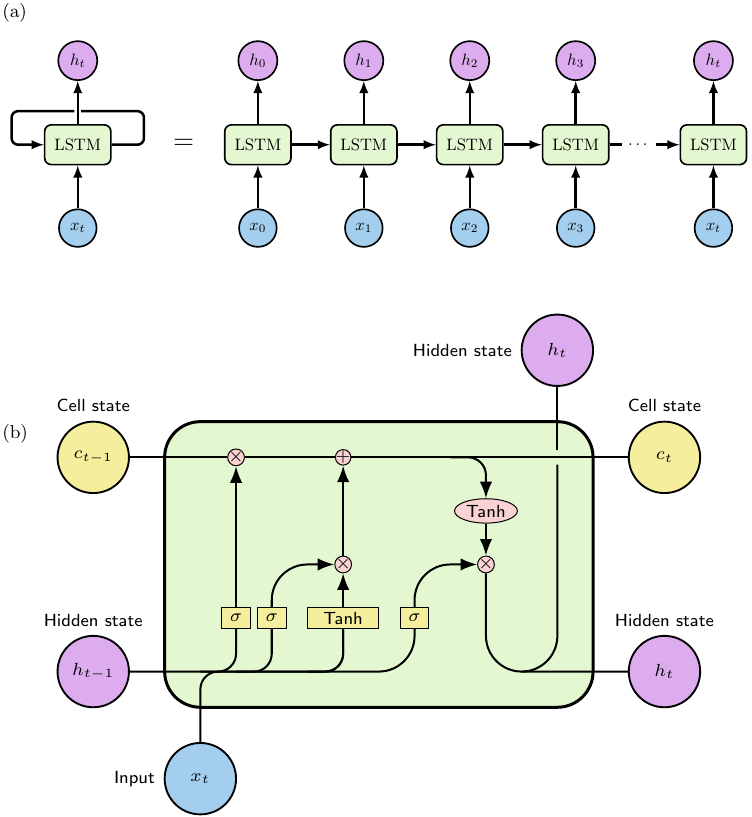}
		\caption{Schematic representations of Recurrent Neural Network (RNN) and Long Short-Term Memory (LSTM) architectures. (a) Depicts the structure of an RNN, characterized by a sequence of repeating modules. (b) Illustrates the intricate architecture of an LSTM, which is a specialized form of RNN designed to mitigate the vanishing gradient problem inherent in traditional RNNs.
		}
		\label{fig:cell}

	\end{figure}

	Recurrent neural network (RNN) provide a solution to time series data problem. These networks are characterized by their looping connections, enabling the retention and transfer of information over time. As shown in Fig.~\ref{fig:cell}(a), a segment of the neural network, examines an input $x_t$ and generates an output $h_t$. This loop structure allows information to propagate from one stage of the network to the subsequent stage. Thus, an RNN can be conceptualized as multiple duplicates of the same network, each relaying a message to its subsequent copy. This design allows RNNs to maintain a form of 'memory' over the input data, a feature particularly useful for tasks involving sequential information.
	
	LSTM (Long Short-Term Memory) is a type of RNN architecture that was introduced by Hochreiter and Schmidhuber~\cite{Hochreiter1997} to solve the problem of vanishing and exploding gradients in standard RNNs. LSTMs are designed to remember and forget information over long sequences, making them particularly effective for applications involving time-series data.
	
	The LSTM architecture comprises four main components: the cell state, the input gate, the forget gate, and the output gate (Fig.~\ref{fig:cell}(b)). These work in unison to regulate the flow of information through the LSTM.
	
	The LSTM update equations are as follows:
	
	\begin{align}
		f_t &= \sigma(W_f \cdot [h_{t-1}, x_t] + b_f), \\
		i_t &= \sigma(W_i \cdot [h_{t-1}, x_t] + b_i), \\
		\tilde{C}t &= \tanh(W_C \cdot [h{t-1}, x_t] + b_C), \\
		C_t &= f_t * C_{t-1} + i_t * \tilde{C}t, \\
		o_t &= \sigma(W_o \cdot [h{t-1}, x_t] + b_o), \\
		h_t &= o_t * \tanh(C_t).
	\end{align}
	
	where:
	\begin{itemize}
		\item $x_t$ is the input at time step $t$.
		\item $h_t$ is the hidden state at time step $t$.
		\item $C_t$ is the cell state at time step $t$.
		\item $f_t$ is the forget gate's activation vector at time step $t$.
		\item $i_t$ is the input/update gate's activation vector at time step $t$.
		\item $o_t$ is the output gate's activation vector at time step $t$.
		\item $\tilde{C}_t$ is the candidate cell state created by the input/update gate at time step $t$.
		\item $\sigma$ is the sigmoid activation function.
		\item $*$ denotes element-wise multiplication.
		\item $W$ and $b$ are the learnable parameters of the LSTM cell.
	\end{itemize}
	
	The LSTM operates by first deciding what information to discard from the cell state, as controlled by the forget gate ($f_t$). The input gate ($i_t$) and the creation of a new candidate cell state ($\tilde{C}_t$) decide what new information to store in the cell state. The cell state is then updated to $C_t$ based on these decisions. Finally, the output gate ($o_t$) decides what part of the cell state is going to be outputted. This output, $h_t$, is then passed to the next time step.
	
	In summary, the LSTM effectively tackles the challenge of learning long-term dependencies in sequence data by controlling the flow of information using a system of gating units. This makes LSTMs a powerful tool for many applications in time-series analysis and natural language processing.

	\section{Training strategy}
	During training (as illustrated in Fig.~\ref{fig:learning}), we input all parameters into the Neural Network (NN) that fully identify the Hamiltonian of the model and the entire driving trajectory $B(t)$. This information enables the prediction of the quantum many-body system's dynamics or the expectation values of a set of local observables $O(t)$ for learning the quantum Hamiltonian. Throughout this process, we always include a description of the initial state.
	
	In our examples, we selected product states as the initial states, denoted as $\rho(0)$. In these scenarios, it is sufficient to provide the expectation values of a collection of local observables. The NN output is either the complete evolution of the expectation values of a selected subset of observables ($O(t)$) for predicting the dynamics of the quantum many-body system, or a driving trajectory $B(t)$ for learning the quantum Hamiltonian. Specifically, we limit our focus to local observables of single and two-qubit.
	
	We use the results of an exact simulation as the provided data. However, future applications could use data from experiments. This approach simplifies the data acquisition process, as generating training data only requires a unidirectional generation process starting from state evolution.
	
	The loss function for the training data is given by the Mean Squared Error (MSE) between the predicted value $\mathbf{Y}^{\mathrm{pred}}$ and the actual value $\mathbf{Y}^{\mathrm{true}}$ of quantum dynamic evolution. This can be mathematically expressed as:
	
	\begin{equation}
		L=\frac{1}{M} \sum_{m=1}^M\left(\mathbf{Y}_m^{\mathrm{true}}-\mathbf{Y}_m^{\mathrm{pred}}\right)^2.
	\end{equation}
	
	Here, $M$ represents the data size. We generate a substantial number of training samples and employ the Adam optimization algorithm to minimize this loss function.
	
	Subsequently, to validate the practicality of our model, we focus on a particular quantum many-body system Hamiltonian, incorporating prior physical knowledge of the system, and train our neural networks to learn these characteristics.
	
	\section{Quantum dynamics prediction and Hamiltonian learning problem}
	This section provides an introduction to the primary setting of the quantum dynamics prediction and Hamiltonian learning problem, along with some fundamental concepts.
	
	We consider an $n$-qubit Hermitian Hamiltonian $H$ of dimension $d\times d$ (where $d = 2^n$), which can be expressed in terms of the Pauli operator basis as follows:
	
	\begin{equation}
		H=\sum_{a=1}^m \lambda_a E_a.
		\label{eq:H}
	\end{equation}
	
	Here, $\lambda_a$ represents real coefficients and for a $k$-local Hamiltonian, the set $\{E_1, \ldots, E_m\}$ is a subset of local Pauli operators acting non-trivially on at most $k$ particles, with $m=O(p o l y(n))$.
	
	In the realm of quantum dynamics prediction, the Hamiltonian model mentioned above is used for a unitary evolution that is defined by a small number of tunable control parameters. Any unitary evolution $U(t,t_0)$ from time $t_0$ to $t$ can be represented~\cite{Breuer2007} through a generating Hamiltonian $H$: $U\left(t, t_0\right)=\mathcal{T}_{\leftarrow}\left[\exp \left(-\frac{2 \pi i}{h} \int_{t_0}^t H\left(t^{\prime}\right) d t^{\prime}\right)\right]$. In this context, $\mathcal{T}_{\leftarrow}$ indicates time-ordering, which simplifies to $U\left(t, t_0\right)=\exp \left(-\frac{2 \pi i}{h} H(t-t_0)\right)$ when $H$ is constant. Our objective is to use the measured values of a series of states $\{O(0) \dots O(t_0)\}$ at the time interval $(0,t_0)$ to predict the value at an arbitrary time $t (t>t_0)$, hence revealing the evolution $U\left(t\right)$.
	
	Reconstructing the Hamiltonian from quantum dynamics is also crucial for quantum sensing and metrology. Contrary to traditional Hamiltonian learning methods, which prepare the eigenstate or thermal state of an arbitrary Hamiltonian, a more intuitive approach is to consider the time evolution of the Hamiltonian, which can be accessed for simulation and experiments. For a $d$-dimensional system, the task then becomes learning the suitable $d\times d$-matrix $H(t)$. As a result, the Hamiltonian learning problem is treated as the reconstruction of the decomposition parameters $\lambda_a(t)$. For a realistic system with certain types of interactions, we can assume a specific Hamiltonian structure. Therefore, for the Hamiltonian in Eq.~\ref{eq:H}, only a polynomial number of parameters are required to characterize the entire system. Additionally, a single eigenstate of such a Hamiltonian can generally encode the system's information~\cite{Qi2019determininglocal,PhysRevX.8.031029,PhysRevB.98.081113}. For example, the time-evolved state for an initial state $\rho(0)$ is represented as a unitary under the Schr\"{o}dinger equation:
	\begin{equation}
		\mathcal{H}_t(\rho)=U(t) \rho(0) U(t)^{\dagger},
	\end{equation}
	
	where $U(t)=e^{-iHt}$. Here we use $\mathcal{H}_{t}$ to denote the Hamiltonian evolution channel with a time interval $(0,t)$. If the time is sufficiently short, we can approximate the Hamiltonian evolution by expanding the series and truncating it after a few terms, as follows:
	\begin{equation}
		\begin{aligned}
			\mathcal{H}_t(\rho(0)) = &\rho(0)
			+ i t \sum{a} \lambda_a \left(\rho(0) E_a - E_a \rho(0)\right) + t^2 \sum_{a, b} \lambda_a \lambda_b \left[E_a \rho(0) E_b
			- \frac{1}{2}\left(E_a E_b\rho(0) + \rho(0) E_a E_b\right)\right] \\
			&+ O\left(t^3\right).
		\end{aligned}
		\label{eq:hl}
	\end{equation}
	
	It's worth noting that the first- and second-order terms of the evolution channel store the decomposition parameters of the Hamiltonian operator accurately, providing an opportunity to extract the Hamiltonian information from the dynamics.
	
	We can consider a plausible learning path for the model in both quantum dynamics prediction and Hamiltonian learning problems. Initially, the model receives the values of local observables $\{O(0) \dots O(t_0)\}$ of the quantum state $\{\rho(0)\dots \rho(t_0)\}$ and the Hamiltonian parameter $\{\lambda_a(0) \dots \lambda_a(t_0)\}$. It then maps these to the evolution of the quantum state $\rho(t)$ and Hamiltonian $H(t)$. Afterward, it deduces the evolution operator $U(t)$ from the data. Finally, using the obtained evolution operator $U(t)$, it provides the observables of the quantum state or Hamiltonian parameter at a later time. It's clear that a model that performs effectively for both problems has managed to learn a representation of the evolution of a quantum many-body system.
	
	\section{Driving filed for Hamiltonian}
	We train the Neural Network on arbitrary time-dependent trajectories, $B(t)$, sampled from a broad class of functions, enabling the NN to later predict the dynamics of expectation values, $O(t)$, for unseen time-dependent driving trajectories. We found that sampling these trajectories from a random Gaussian process~\cite{Liu2019} was particularly effective. This approach, by offering flexibility in data fitting through control of mean and variance, becomes a fundamental choice in statistics. Specifically, we employ a mixture of Gaussian processes with randomly varied correlation times. This approach enables the NN to learn essential features for predicting dynamics for various time-dependent functions.
	
	\subsection{Generating Random field from Gaussian random process}
	There are numerous methods available for generating Gaussian random functions~\cite{Liu2019}. In this section, we delineate the specific method we employ. We begin by discretizing the time interval of interest with a time step of $\triangle t$ (in our case, we set $\triangle t =0.1$). This allows us to define a vector $\vec{d}=(D(0), D(\Delta t), D(2 \Delta t), \ldots)^T$, encapsulating the trajectory of the time-dependent parameter.
	
	Next, we construct the correlation matrix $C$ with elements $C_{n m}=\left\langle d_n d_m\right\rangle=$ $\left.c_0 \exp \left[-(n-m)^2 \Delta t^2 / 2 \sigma^2\right)\right]$, where we utilize a Gaussian correlation function with a correlation time $\sigma$ (although other functional forms could also be employed).
	
	Given that the correlation matrix is real and symmetric, it can be diagonalized as $C=Q \Lambda Q^T$, where $\Lambda$ is a diagonal matrix containing the eigenvalues and $Q$ is an orthogonal matrix. Consequently, we can generate the random parameter trajectory as $d=Q\sqrt{\Lambda}\vec{x}$, where $\vec{x}$ is a vector with independent random variables drawn from a unit-width normal distribution ($\left\langle x_n\right\rangle=0$ and $\left\langle x_n x_m\right\rangle=\delta_{nm}$).
	
	While we could theoretically sample using Fourier series with random coefficients, doing so would generate periodic functions and potentially introduce unwarranted correlations between the values of the drive at early and late times. To avoid this, we chose not to use this method.
	
	To ensure the generated random functions are representative of a broad array of arbitrary time-dependent functions, we randomly select the correlation amplitude $c_0$ and time $\sigma$ for each training trajectory. Specifically, $c_0$ is chosen uniformly from the range $c_0\in[0,4]$, correlating to a maximum magnetic field $\abs{B}$ of approximately 5, and $\sigma$ is chosen uniformly from the wider interval $\sigma\in[1,9]$.
	\subsection{Generating other time-dependent fields for evaluating}
	As elucidated in the main text, evaluating the performance of our model and demonstrating its capability to predict dynamics for a variety of driving trajectory functional forms necessitates testing the model on random periodic driving fields and different types of quenches.
	
	For the periodic drivings, we generate 100 functions of the form $B(t) = A\cos(\omega t)$, where the amplitudes $A$ and frequencies $\omega$ are chosen randomly from intervals $[-3,3]$ and $[0.1,4]$, respectively.
	
	Regarding quenches, we create 100 step functions. The heights of the steps are randomly selected from the interval $[-3,3]$ and the quenches occur at random times. This diverse testing set is intended to provide rigorous testing of the model's performance across different types of driving trajectories.
	
	\section{Study of 1D TFIM}
	
	In order to measure the effectiveness of our approach in forecasting the behavior of a driven quantum many-body system, we focus on one-dimensional (1D) spin models. We analyzed these models under periodic boundary conditions.
	
	The model we studied in this paper was the Transverse Field Ising Model (TFIM), where the spin ring is disturbed from equilibrium by a time-dependent transverse field that is identical across the entire system.
	
	The Hamiltonian for the TFIM in a system of size N is represented as follows:
	
	\begin{equation}
		H_{\text{TFIM}} = -\sum_{i=0}^{N-1}(  J\sigma_i^z \sigma_{i+1}^z+B(t)\sigma_i^x ).
	\end{equation}
	
	Here, $\sigma_i^\alpha$ (where $\alpha$ can be $x, y, z$) are Pauli operators acting on site $i$.
	
	The TFIM is a widely studied foundational model with a multitude of experimental applications. It exhibits complex physics, including a quantum phase transition at $B(t)=J$~\cite{Sachdev2011}, and a diverse range of phenomena when driven from equilibrium, especially during quenches. In our study, we standardize $J = 1$, which merely sets the energy scale.
	
	In our approach, the NN is trained on Gaussian random fields $B(t)$ encompassing a wide range of field amplitudes. Specifically, for the numerical experiments examined in this study, we generated 100,000 random realizations of a Gaussian random process for $B(t)$.
	
	When our system interacts weakly with its environment, its dynamics is governed by the master equation. We solve both the Schr\"{o}dinger equation and the master equation for the 1D TFIM driven out of equilibrium by these random transverse fields for a specific system size $N$ of interest. The initial state is prepared in an arbitrary product state, represented as $\otimes_i (\sqrt{p}|0\rangle + \sqrt{1-p}|1\rangle)$, where $p$ is a variable that ranges from 0 to 1.
	
	We then compute the evolution of a selected subset of observables. It's important to note that, for all the spin models we investigate in this work, we consistently choose the local spin operators $\langle \sigma_i^\alpha \rangle$ , along with correlators of the type $\langle \sigma_i^\alpha \sigma_{i+l}^\beta \rangle$ with $\alpha, \beta=x, y, z $ as our observables.
	
	However, this selection is purely illustrative and does not imply a requirement to train the NN over all these observables for proper learning of the dynamics. Indeed, other studies have confirmed that successful training of the NN can be achieved with fewer observables of interest.
	
	To assess the performance of the fully trained NN and demonstrate its ability to accurately predict dynamics for any type of time-dependent field, we test the NN on a new set of Gaussian random fields (ones not encountered during training), various types of quenches, and random periodic driving fields.
	
	For periodic drivings, we generate a set of functions in the form $B(t) = A\cos(\omega t)$, where $A$ and $\omega$ are randomly chosen values. For quenches, we generate a set of step functions, randomly selecting the heights of steps and the times at which the quenches occur. Additional details regarding the range of parameters chosen for generating our periodic and quench trajectories can be found in the Appendix.
	
	\subsection{Numerical Results}
	We proceed to apply our model to the tasks of quantum state dynamics prediction and Hamiltonian learning under various conditions, demonstrating the model's power and flexibility. These conditions involve different driving fields for our system, as well as quantum many-body systems under the influence of decoherence effects or operating within a noiseless environment.
	
	\begin{figure}[!htbp]
		\centering
		\includegraphics[width=1\textwidth]{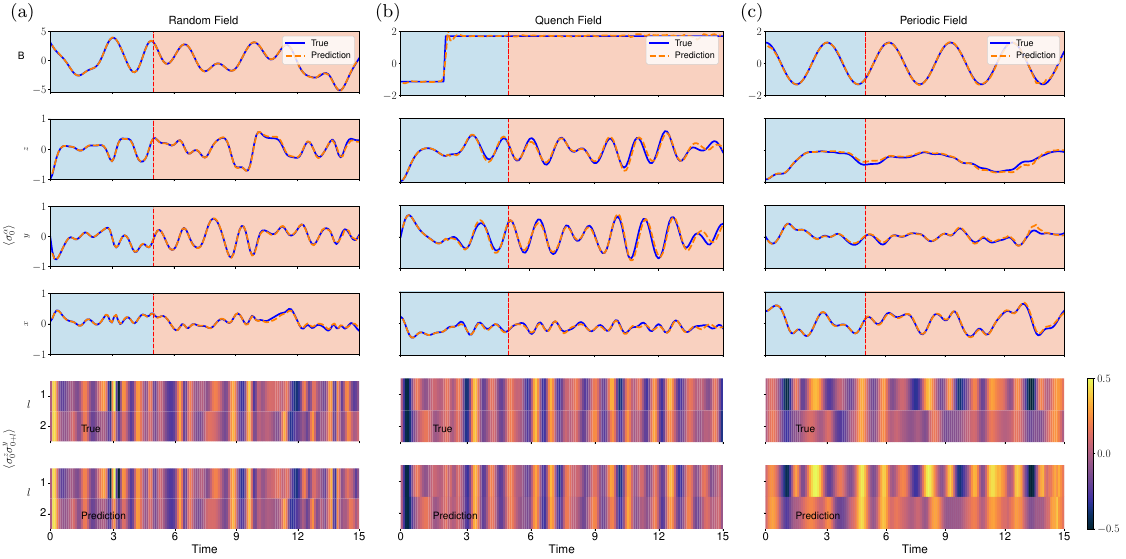}
		\caption{Demonstrating the results of our model in predicting the noiseless dynamics of a spin system under arbitrary driving. The model is trained on Gaussian random driving fields within a fixed time interval ($t\in [0,5]$), and the model's predictive performance is then evaluated for various driving fields and a larger time interval ($t\in [0,15]$). Here, the actual and predicted evolution of the external driving field $B$ (displayed in the top row) and observables $\langle \sigma_0^\alpha \rangle$ (shown in the middle three rows) and $\langle \sigma_0^z \sigma_{0+l}^y \rangle$ (presented in the bottom two rows) are compared for a transverse-field Ising spin ring of size $N=5$. The comparisons are shown for different classes of driving fields: (a) an unseen random field, (b) a quench, and (c) periodic driving. For both random field driving cases, the spins are initialized in a randomly directed, translationally-invariant product state. The time window ($t\in [5,15]$) highlighted in light orange represents prediction beyond the training interval.
		}
		\label{fig:numresults}
	\end{figure}
	
	Firstly, we investigate the performance of our model for a noiseless environment where a 1D TFIM of size $N=5$ was considered. Fig. \ref{fig:numresults} demonstrates the consistency between model prediction and ground truth in three examples with: (a) an unseen random driving field, (b) a quench field and (c) a periodic driving field. In this dual-purpose approach, the expected values of the selected local observables $\langle \sigma_0^\alpha \rangle$ and $\langle \sigma_0^\alpha \sigma_{0+l}^\beta \rangle$ are predicted based on the driving
	trajectory of the external field $B(t)$, while the external driving field is predicted based on dynamics trajectories of the selected local observables. Our model is trained within the fixed time interval ($t\in [0,5]$). As shown in Fig. \ref{fig:numresults}, even for a larger time interval ($t\in [0,15]$), our model is able to provide an appreciable level of accuracy in both predicting quantum state evolution and learning Hamiltonian.
	
\begin{figure}[!htbp]
		\centering
		\includegraphics[width=1\textwidth]{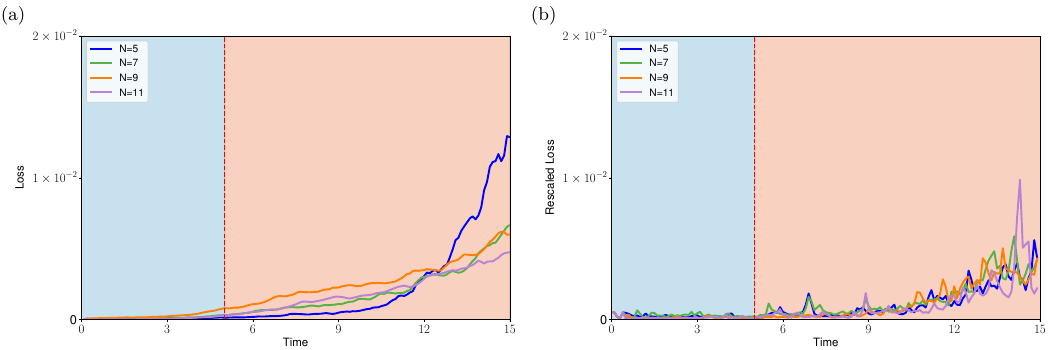}
		\caption{The noiseless evolution of our model's prediction loss, measured by Mean Squared Error (MSE), averaged over 100 realizations of Gaussian random fields with different system size $N=5,7,9,11$. Two aspects are considered: (a) all selected observables $\langle \sigma_0^\alpha \rangle$ and $\langle \sigma_0^\alpha \sigma_{0+l}^\beta \rangle$, and (b) the external driving field $B$. Note that the loss related to the external driving fields is rescaled by a factor of 5, due to the field range being approximately $[-5,5]$.}
		\label{fig:losses}
	\end{figure}

	A thorough evaluation of our models is performed by training them on noiseless evolutions of systems of varying sizes ($N=5, 7, 9, 11$). Subsequently, we assess the MSE losses between the model predictions and the actual system values. As shown in Fig.~\ref{fig:losses}(a), the MSE losses for all chosen observables remain relatively low (below $10^{-3}$) throughout the training time interval ($t\in [0,5]$). During the extrapolation time interval ($t\in (5,15]$), a noticeable increase in losses is observed, peaking at approximately $10^{-2}$. However, the overall losses remain relatively low, roughly around $10^{-2}$.
	
	For the prediction of driving fields, Fig.~\ref{fig:losses}(b) illustrates somewhat larger fluctuations in MSE losses for all cases. Considering the field range in these cases extends from $-5$ to $5$, we have rescaled the loss values by a factor of $5$. The rescaled losses remain below $10^{-3}$ in the training time interval, increasing during the extrapolation time interval, but never exceeding $10^{-2}$ at any point. These results underscore the robustness of our model in learning the dynamics of quantum many-body systems of different sizes.
	
	Secondly, the total time required to measure the temporal records may reach or even surpass the coherence time of the experimental apparatus. As a result, the gathered temporal records include the decoherence effect, which can negatively impact the prediction accuracy. To address this issue, we also conduct a numerical study on the performance of our model under the influence of decoherence.
	
\begin{figure}[!htbp]
		\centering
		\includegraphics[width=0.48\textwidth]{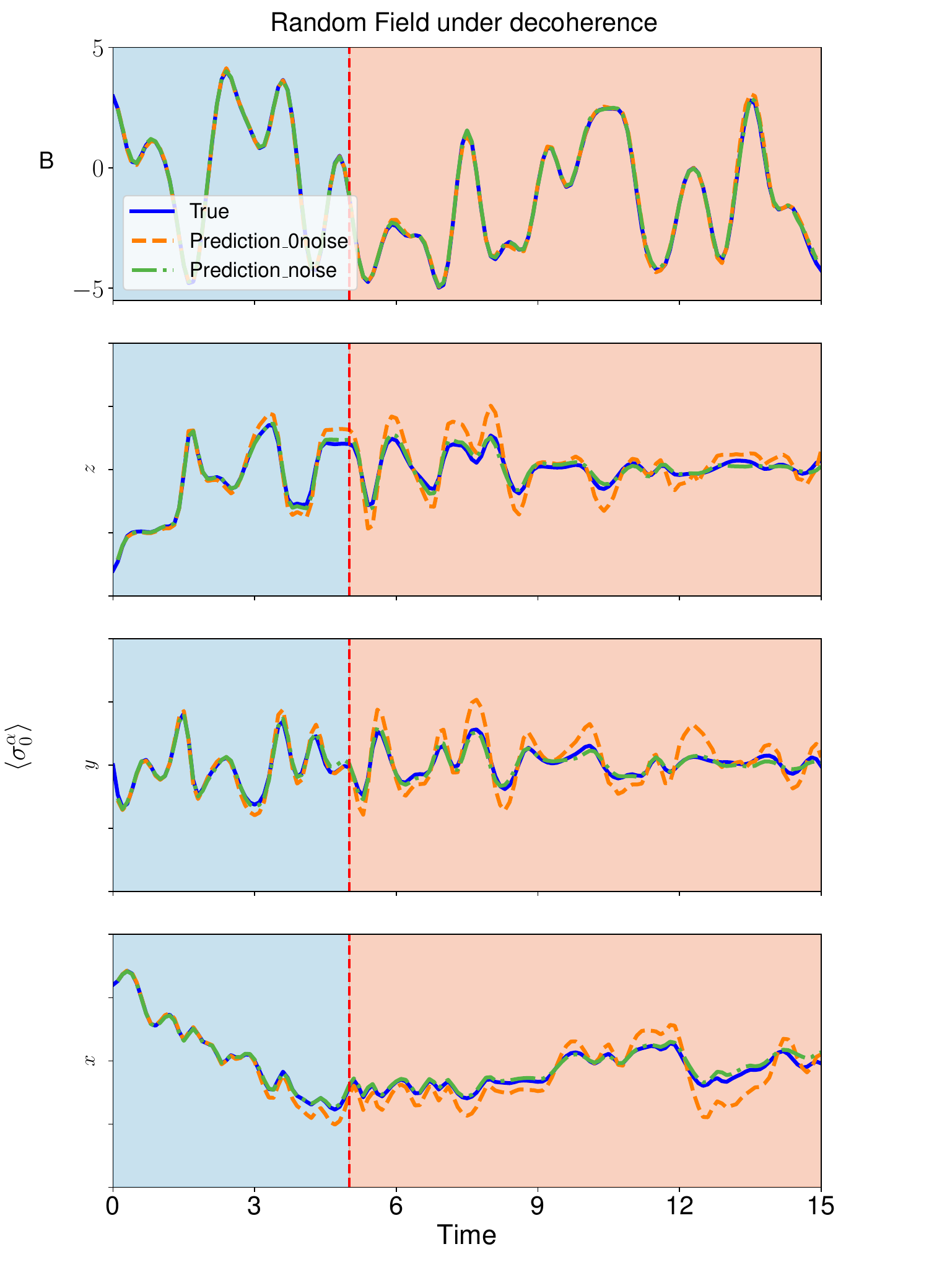}
		\caption{The performance of our model in predicting the decoherence dynamics of a spin system under random field driving. Specifically, we depict the predicted evolution of the external driving field $B$ and local observables $\langle \sigma_0^\alpha \rangle$ using the trained models, denoted as `Prediction\underline{\hspace{0.5em}}0noise' and `Prediction\underline{\hspace{0.5em}}noise'.}
		\label{fig:deco}
	\end{figure}

	In a similar fashion, two distinct models (`Prediction\underline{\hspace{0.5em}}0noise' and `Prediction\underline{\hspace{0.5em}}noise') are trained using both noiseless and decoherence data (additional details are available in the Appendix). These models are subsequently used to learn the expectations values of the local observables ${ \sigma_0^x, \sigma_0^y, \sigma_0^z }$ and the driving field $B$, utilizing the decoherence test data. As shown in Fig.~\ref{fig:deco}, the `Prediction\underline{\hspace{0.5em}}noise' model, trained on decoherence data, exhibits superior accuracy when compared to the `Prediction\underline{\hspace{0.5em}}0noise' model. This disparity in accuracy becomes more pronounced over time, particularly during the extrapolation time interval. Nevertheless, both noiseless and decoherence cases show a significant degree of concurrence between estimated and actual trajectories for the driving field.
	
	\section{Details of simulation quantum many-body system dynamics}
	In this study, our objective is to evaluate the effectiveness of our proposed methodology in forecasting the dynamics of quantum many-body systems subjected to external driving. Our focus is specifically centered on 1D spin models under periodic boundary conditions. The evolution of these quantum systems is determined by solving the Schr\"{o}dinger equation in noiseless environments and the master equation in environments subject to decoherence, both of which govern the time evolution of quantum systems.
	
	The Schr\"{o}dinger equation for our system of interest is expressed as:
	
	\begin{equation}
		i \hbar \frac{d|\psi(t)\rangle}{dt} = H |\psi(t)\rangle.
	\end{equation}
	
	In this equation, $|\psi(t)\rangle$ represents the quantum state of the system at time $t$, and $\hbar$ denotes the reduced Planck's constant. In our specific scenario, we simplify our calculations by setting $\hbar=1$.
	
	The master equation, also known as the Lindblad equation, describes how the density matrix $\rho$ of a quantum system evolves over time when it is part of an open quantum system. In general, the Lindblad equation is written as:
	
	\begin{equation}
		\frac{d\rho}{dt} = -i[H, \rho] + \sum_i \sqrt{\gamma_i} \left( L_i \rho L_i^\dagger - \frac{1}{2} {L_i^\dagger L_i, \rho} \right).
	\end{equation}
	
	Here, $H$ is the Hamiltonian of the system, $L_i$ are the Lindblad operators (or collapse operators) that describe the system's interaction with the environment, and $\gamma_i$ are the decay rates associated with each Lindblad operator.
	
	The collapse operators $L_i$ are constructed as follows:
	
	\begin{equation}
		L_i = \sum_{i=0}^{N-1} \sigma_{x}(i).
	\end{equation}
	
	Here, $\sigma_x(i)$ is the Pauli-X operator (also known as the bit-flip operator) perform at the $i$-it qubit. The index $i$ runs over the $N$ spins in the system.
	
	\section{Description of NMR Experiments}
	The experiments were conducted on Gemini, a desktop nuclear magnetic resonance (NMR) quantum computer~\cite{hou2021spinq}.
	Shown in Fig.~2 of manuscript, the two qubits are represented by two connected $^1$H and $^{31}$P in Dimethylphosphite ((CH$_3$O)$_2$PH) molecules.
	$T_1$ and $T_2$ represent the longitudinal and transverse relaxation times, respectively.
	The free evolution of this $2$-qubit system is primarily governed by the internal Hamiltonian, 
	\begin{eqnarray}\label{NMR_Hamiltonian}
		\mathcal{H}_{int}= \pi \nu _1  \sigma_z^1+\pi \nu _2  \sigma_z^2  + \frac{\pi}{2} J_0 \sigma_z^1 \sigma_z^2, 
	\end{eqnarray}
	where $\nu_1$  and $\nu_2$ can be adjusted to $0$ Hz in a rotating frame, and $J_0 = 697.4$ Hz denotes the resonance frequency of the $J$-coupling strength between the spins. To control the system's evolution, transverse radio-frequency (r.f.) pulses serve as the control field, expressed as,
	\begin{eqnarray}
		\label{NMR_RFHamiltonian}
		\mathcal{H}_{rf}=-\frac{1}{2} \omega_1\sum_{i=1}^2 (\cos(\omega_{rf} t+\phi)\sigma_x^{i}+\sin(\omega_{rf} t+\phi)\sigma_y^i).
	\end{eqnarray}
	By adjusting the parameters in the r.f. field (Eq.~\ref{NMR_RFHamiltonian}), such as intensity $\omega_1$, phase $\phi$, frequency $\omega_{\text{rf}}$, and duration, the theoretical achievement of two-qubit universal quantum gates is possible through the combination of the system's internal dynamics(Eq.~\ref{NMR_Hamiltonian})~\cite{lu2017enhancing}.

	\section{Description of Superconducting qubit Experiments}
	
\begin{figure}[!h]
		\begin{center}
			\includegraphics[width=0.5\textwidth]{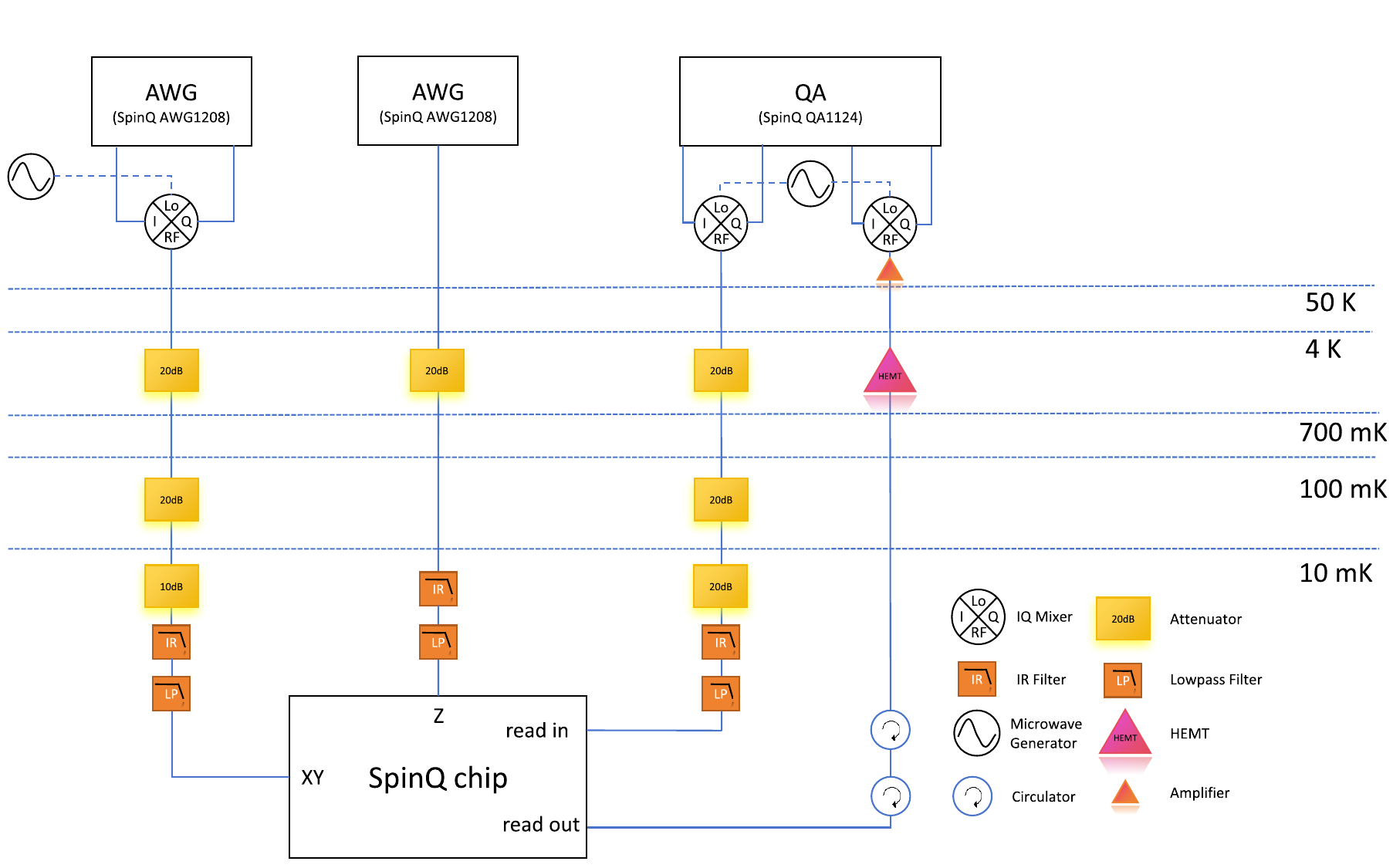}
			\caption{Diagram of control electronics and wiring of a superconducting-qubit system.}
			\label{sp:fig1}
		\end{center}
	\end{figure}

	The superconducting quantum computer used in this experiment consists of a low-temperature part and a room-temperature part. 
	Here we give the schematic diagram of control electronics and wiring in Fig.~\ref{sp:fig1}. 
	The chip is 2D ladder with 20 transmon qubits. The qubits operating at frequencies of $4.95$ GHz and $5.45$ GHz were employed  in our experiments, denoted as $Q_1$ and $Q_2$ respectively. As depicted in Fig.~2(b) of manuscript, $Q_1$ has a T1 time of $28.72~\mu s$ and a T2 time of $5.20~\mu s$, while $Q_2$ has a T1 time of $22.86~\mu s$ and a T2 time of $2.92~\mu s$. 
	
	\subsection{Pulses to drive and read qubits}
	XY control signals, which drive the qubit transitions along the XY plane, are generated by modulating microwave signals through a mixer connected to an Arbitrary Waveform Generator. 
	Z control signals, which are responsible for evolving the qubit along the Z direction, are generated directly by a single channel of the Arbitrary Waveform Generator. 
	This channel is equipped with both dc Z bias and ac Z bias output capabilities, allowing it to perform both slow and fast Z bias control.
	After undergoing multi-layer attenuation and filtering at low temperatures, these signals are delivered to the qubits.
	
	The readout signal is generated by the Arbitrary Waveform Generator through a mixer that modulates a microwave signal. After multi-layer attenuation and filtering, the signal is directed to the readout resonator coupled with the qubits.
	The signal then passes through the readout resonator and through multiple cascaded circulators. It is then amplified by low-temperature amplifiers, followed by further amplification by room-temperature amplifiers to prepare the signal for detection. Once amplified, the signal undergoes demodulation, and is finally collected and processed by the ADC integrated on the control board.
	
	\subsection{Calibrations}
	In this section, we will introduce our calibration of quantum gates, amplitude and phases of the square-shaped wave.
	
	\subsubsection{Calibration of Quantum gates}
	The gate used in the experiment is a $20$~ns microwave pulse with a Gaussian envelope. The amplitude is calibrated by Rabi oscillations.
	Additionally, the protocol is known as the Derivative Reduction by Adiabatic Gate (DRAG)~\cite{motzoi2009simple} is incorporated to correct for errors due to the presence of the $\ket{2}$ state.

	\subsubsection{Calibration of Square-shaped Wave}
	Due to interactions such as reflection, dispersion, and the effects of parasitic capacitance and inductance in the line, the square-shaped wave applied in Z bias control line may become distorted.These distortions can significantly impact the fidelity and accuracy of quantum operations. Therefore,it is essential to calibrate the square-shaped wave by measuring its transfer function~\cite{neill2018blueprint}.
	
	Ideally, after the application of the square-shaped wave, the quantum bits should not accumulate any phase. However, in actual circuits, the square-shaped wave can become distorted due to convolution with the line's transfer function, leading to residual phases after the application of the square-shaped wave.
	To mitigate this issue, it's necessary to identify the transfer function and perform a deconvolution with the desired square-shaped wave to accurately shape our input signal.
	
	This calibration process is most effectively conducted in the frequency domain, where the convolution operations of the time domain translate into simpler multiplication operations. By shifting to the frequency domain, we can apply the following mathematical model,
	\begin{eqnarray}
		H(\omega) = 1 + \sum_{n=1}^{k} \frac{i \alpha_n \omega}{i \omega + \tau_n}.
	\end{eqnarray}
	which describes multiple overshoots (positive $\alpha_n$) or undershoots (negative $\alpha_n$) which exponentially decay to the final value with time constants $\tau_n$.
	
\begin{figure}[!htbp]
		\begin{center}
			\includegraphics[width=0.4\textwidth]{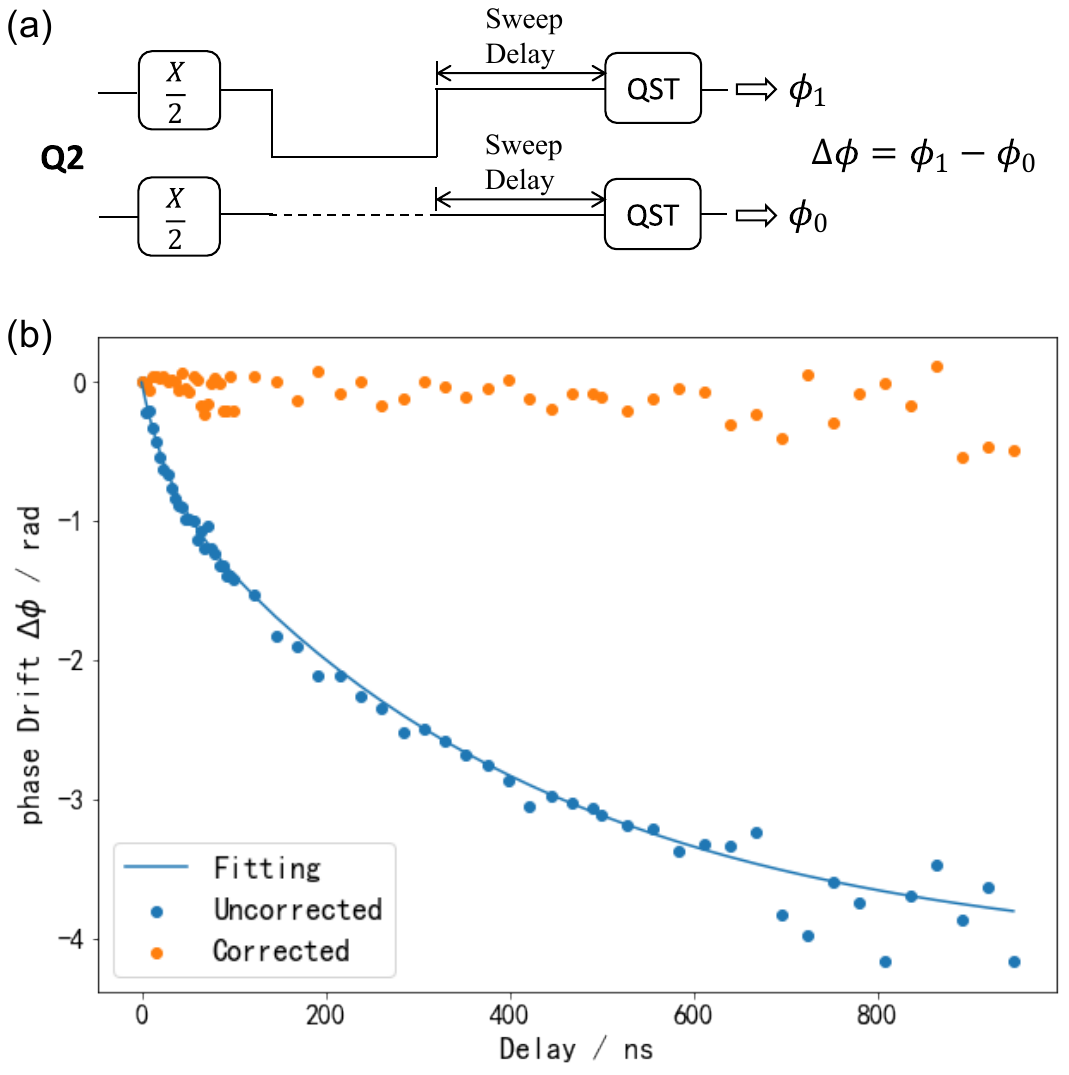}
			\caption{Measurement of the Z-line transfer function for Q2.(a) Pulse diagram: first prepare Q2 in a superposition state, then apply and not apply the square-shaped wave to Q2, change the time after the square-shaped wave (Delay), and measure the phase after the delay using quantum state tomography (QST).(b) Fitting experimental data.}
			\label{sp:fig2}
		\end{center}
	\end{figure}

	The specific experimental procedure is depicted in Fig.~\ref{sp:fig2}

	\subsubsection{Calibration of Amplitude of Square-shaped Wave }
	When adjusting the amplitude of the square-shaped wave applied to $Q_2$, the SWAP frequency between $Q_1$ and $Q_2$ also changes accordingly. The SWAP frequency reaches its minimum when the frequencies of $Q_1$ and $Q_2$ are equal, as described by the equation:
	\begin{eqnarray}
		f_{\mbox{swap}} = \sqrt{4B_0^2 + \Delta^2} 
	\end{eqnarray}
	In this formula, $\Delta$ represents the frequency difference between the two qubits. $B_0$ denotes the coupling strength between them.
	The optimal square-shaped wave amplitude, which minimizes $f_{\mbox{swap}}$, corresponds to the point where $\Delta$ is zero. This is shown in Figure~\ref{sp:fig2}.
	
	\begin{figure*}[!htbp]
		\begin{center}
			\includegraphics[width=0.8\textwidth]{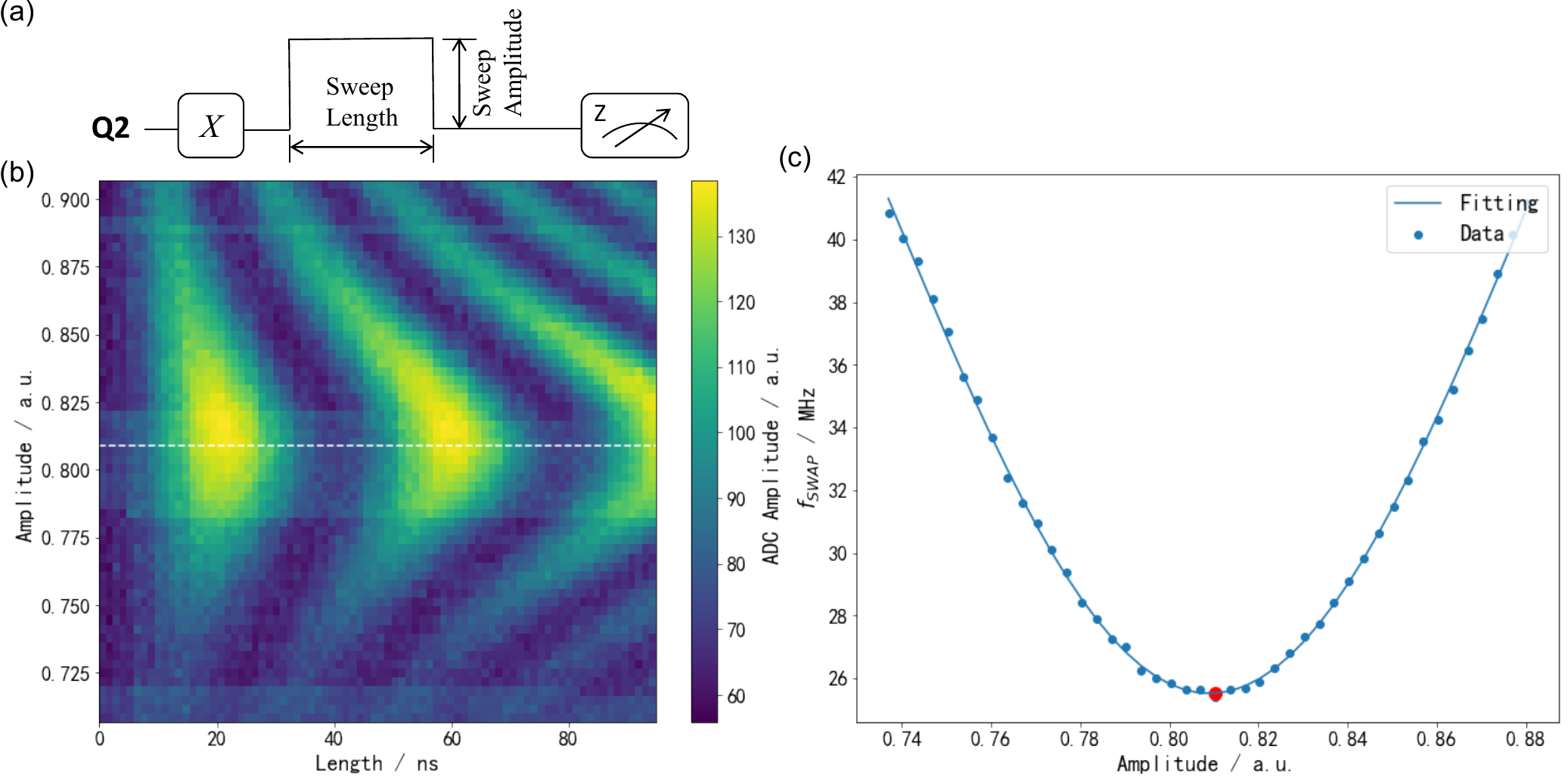}
			\caption{(a) Prepare Q2 in the $\ket{1}$ state, scan the amplitude and duration of the square-shaped wave. SWAP oscillations occur between Q1 and Q2, at which time we measure the state of Q2. (b) Graph showing the relationship between the state of Q2 and the amplitude and duration of the square-shaped wave; the white dashed line indicates that at this point $E_{|01\rangle} = E_{|10\rangle}$. (c) From (b), extract the data for oscillation frequency and square-shaped wave amplitude and perform fitting. The red dots indicate the square-shaped wave amplitude we are looking for.}
			\label{sp:fig3}
		\end{center}
	\end{figure*}
	
	\subsubsection{Compensation of Phase of Square-shaped Wave}
	After calibrating the square-shaped wave, we need to compensate for the extra phase brought by the square-shaped wave. Because during the square-shaped wave application phase, the rotation reference frame of the two qubits undergoes a temporary transformation, accumulating extra phase. To correct this, the extra phase must be experimentally measured. Once measured, a virtual Z rotation can be applied to compensate for the phase changes in $Q_1$ and $Q_2$ post-application of the square-shaped wave~\cite{mckay2017efficient}.
	\begin{figure}[!htbp]
		\begin{center}
			\includegraphics[width=0.4\textwidth]{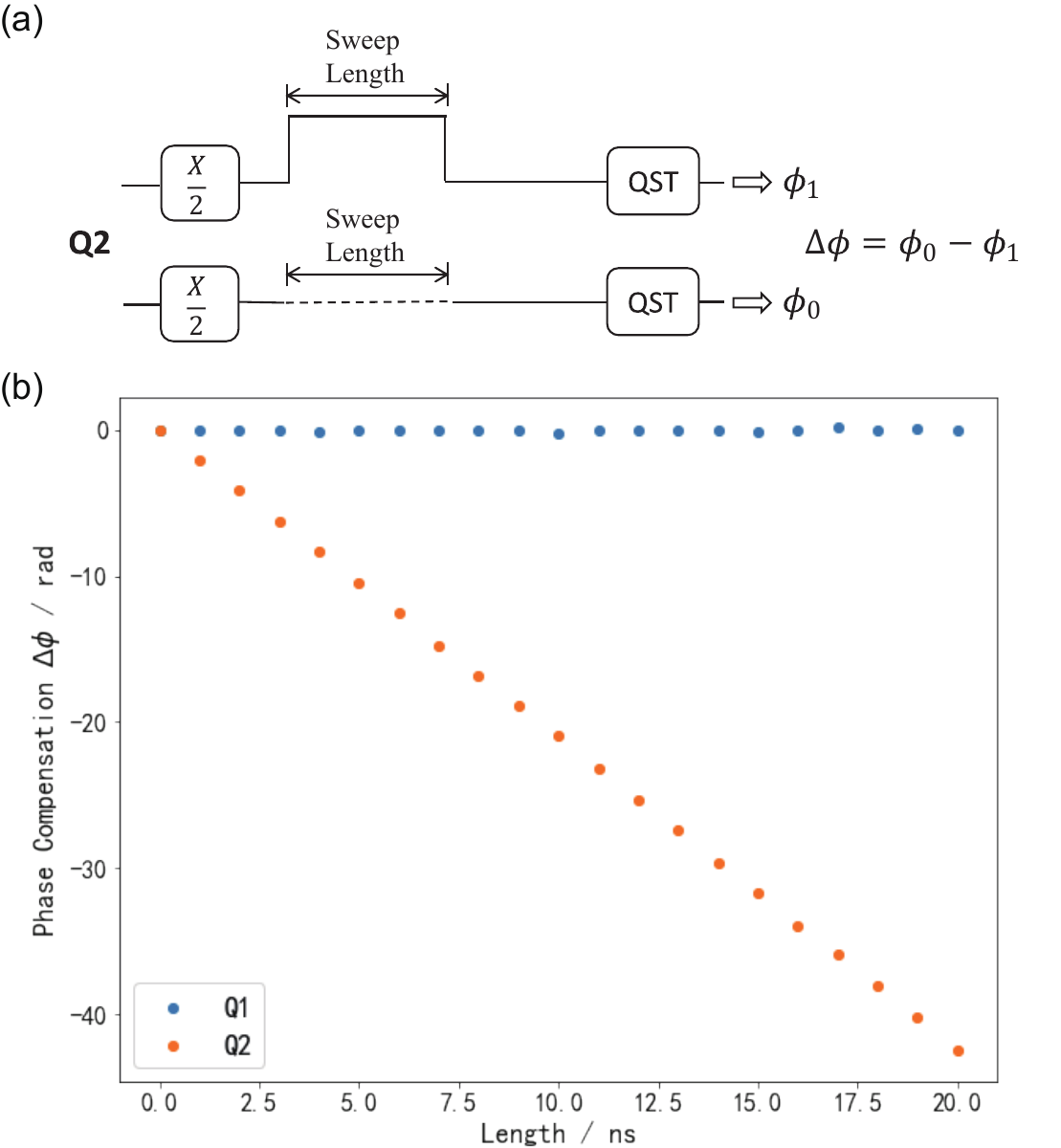}
			\caption{(a) Tune away Q1, prepare Q2 in a superposition state, apply and not apply the square-shaped wave to Q2, then determine the excess phase accumulated by Q2 due to the square-shaped wave. Similarly, swap Q1 and Q2 to determine the excess phase accumulated by Q1 due to the square-shaped wave. (b) Graph showing the relationship between the excess phase accumulated by Q1 and Q2 and the duration of the square-shaped wave.}
			\label{sp:fig4}
		\end{center}
	\end{figure}
	
	\section{Comparative Analysis with Alternative Predictors}
	In this section, we provide an analytical comparison between our method and the RNN predictor, specifically focusing on the aspect of prediction precision.
	
	\begin{figure}[htbp]
		\centering
		\includegraphics[width=0.48\textwidth]{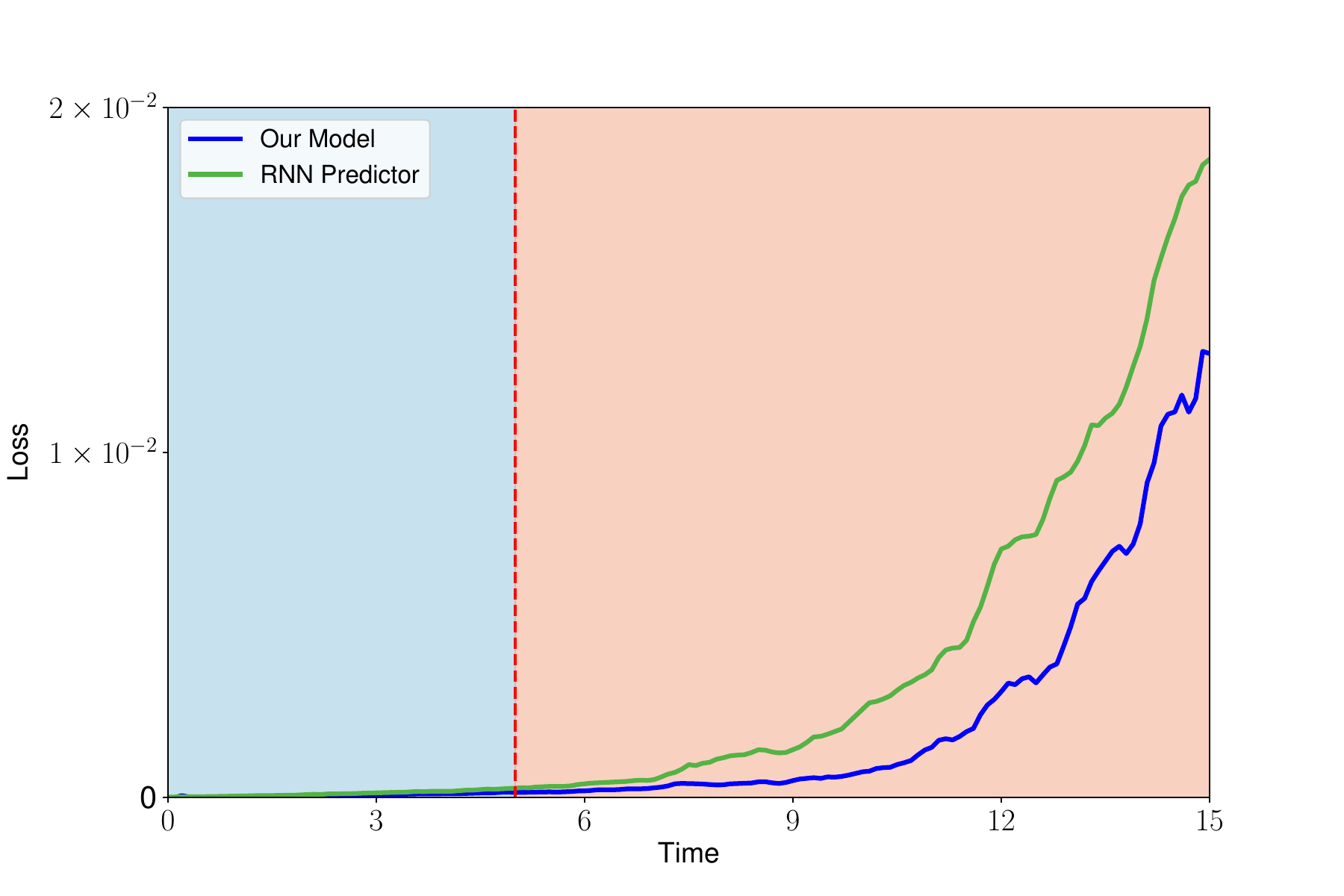}
		\caption{The predictive capabilities of our model and the RNN predictor in modeling the dynamics of an arbitrarily driven spin system. We display the prediction error over time (for a TFIM spin ring of size $N = 5$), driven by 100 random driving fields. The spins are initially arranged in a translationally invariant product state with random spin direction. }
		\label{fig:comp}
	\end{figure}
	
	It's important to note that numerous quantum dynamic predictors have been proposed, with the majority being grounded in machine learning~\cite{PhysRevX.10.011006,Lin2022,Mohseni2022deeplearningof,sornsaeng2023quantum}. For comparison purposes, we've selected the RNN predictor~\cite{Mohseni2022deeplearningof}, which closely aligns with our method. We use the original version of the predictor in this analysis, while our attention is primarily centered on the predictor's performance.
	
	Notably, the predictor can only provide predictions for the values of the observables of the evolving quantum state, not for the Hamiltonian. In addition, the method for Hamiltonian learning can only make predictions within the time range of the training data, which significantly differs from our method. Therefore, our comparison is solely based on the quantum state observables of interest.
	
	As shown in Fig.~\ref{fig:comp}, our method excels by obtaining lower error rates in predicting quantum many-body dynamics. This significant reduction in error is primarily due to our distinctive approach that amalgamates LSTM with an initial state encoder. By exploiting the advantages of these techniques, we can substantially improve the precision and dependability of our predictions. This underscores the viability of our method in a variety of applications where accurate modeling of quantum dynamics is of utmost importance. In addition, unlike other techniques, our method can also be employed for Hamiltonian learning. This makes it a multi-purpose model, a feature that has not been presented in other studies.

\end{document}